# Superstatistics of the screened Kratzer potential with Modified Dirac Delta and Uniform Distributions

P. O. Amadi[1*], C. O. Edet[1], U. S. Okorie[1&2], G. T. Osobonye[3], G. J. Rampho[4] and A. N. Ikot[1&4].

[1]Department of Physics, Theoretical Physics Group, University of Port Harcourt, Choba, Nigeria
[2]Department of Physics, Akwa Ibom State University, Ikot Akpaden, P.M.B. 1167, Uyo. Nigeria.
[3] Department of Physics, Federal College of Education (Technical), Omoku, Rivers State, Nigeria.
[4]Department of Physics, University of South Africa, Florida 1710, Johannesburg, South Africa.

**Abstract**

We solve the Schrodinger equation to obtain the energy eigenvalues expression of the screened Kratzer potential (SKP) model. With the energy eigenvalues, we evaluated for the partition function within the framework of superstatistics and extended to study the thermodynamic function for some selected diatomic molecules including HCl, LiH and $H_2$. The modified Dirac delta and uniform distribution comparatively in each case in the absence and the presence of the deformation parameter were considered.

1. Introduction

The formulation of the Schrodinger equation by Erwin Schrodinger in 1926 has contributed immensely in the study of quantum mechanics. This arises from the various potential models proposed by researcher to study physical problems within the realms of relativistic and non-relativistic quantum mechanics [1-20]. In addition, the solution of the Schrödinger equation has been extended to study to the various diatomic molecules and statistical mechanics of thermodynamic function obtained from partition function [21-24].

Thermodynamic properties have been explored vastly for various quantum system [25-27] and recently the superstatistics have been introduced to study the thermodynamic properties. Superstatistics is the superposition of two different statistics. Superstatistics was proposed by Wilk and Wlodarczyk [28]. Later on, Becks and Cohen [29] laid its formalism. The basic idea that motivated the formulation of superstatistics rest on finding techniques to address the complex physical systems. These complex systems are characterized by their non-equilibrium and stationary state that results from the fluctuation of varying spatiotemporal scales [30]. It is of note that the essential component of superstatistics is the intensive parameter, the fluctuation parameter $\beta$. Before now, investigations have been carried out within the confines of superstatistics [31-37]. To mention a few; Okorie et al. [38] studied for the Dirac Delta and Uniform distribution for Modified Rosen-Morse potential, Sobhani et al. [39] used the harmonic oscillator to study the q-deformed superstatistics and ordinary statistics within the cosmic string, Sargolzaeipo et al. [40] investigated the deformed formalism for the different kinds of distribution, and Sattin [41] investigated the relationship between superstatistics and temperature fluctuation.

In this study, we intend to examine the screened Kratzer potential (SKP) [42] of the deformed formalism and also for uniform distribution, taking into consideration lithium hydride $(LiH)$, hydrogen chloride $(HCl)$ and hydrogen dimer $(H_2)$ diatomic molecules. Thus our work would be presented as follow. Solution of Schrodinger equation of the SKP using factorization method [43] would be presented in section two. Thermodynamic properties for the modified Dirac delta and uniform distribution would be featured in section 3 and 4 respectively. Section 5 will be the application of the diatomic molecules of LiH, HCL and $H_2$, and finally, Section 6 is the conclusion of this study.

## 2. Bound state solution of the Screened Krazter Potential (SKP)

The Radial Schrodinger equation is given as [6],

$$\frac{d^2\psi(r)}{dr^2} + \frac{2\mu}{\hbar^2}\left[E - V(r) - \frac{l(l+1)\hbar^2}{2}\right]\psi(r) = 0 \qquad (1)$$

Where $\mu$ is the reduced mass, $\hbar$ is the Planck constant, and $E$ is the energy of the system. The screened Kratzer potential is defined as [42],

$$V(r) = \left(\frac{A}{r} + \frac{B}{r^2}\right)e^{-\alpha r}, \qquad (2)$$

where $A = -2D_e r_e$ and $B = D_e r_e^2$. $D_e$ is the dissociation energy, $r_e$ is the equilibrium bond length, and $\alpha$ is the potential screening parameter.

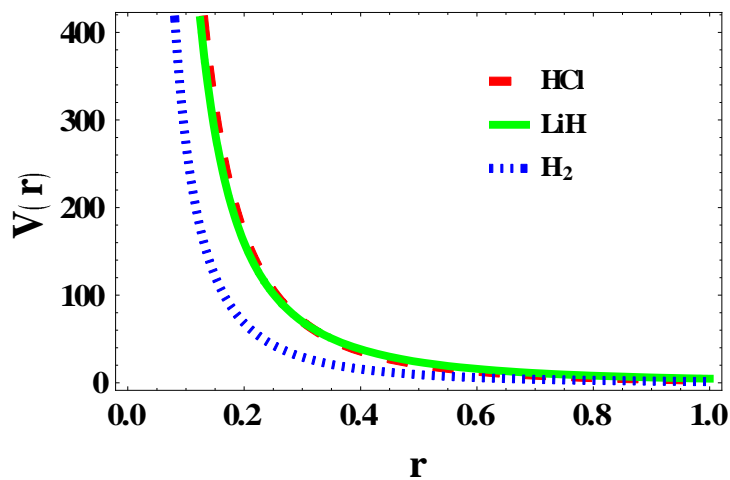

Fig. 1. Behavior of the different diatomic molecules for the SKP.

Substituting eqn (1) into (2) to obtain,

$$\frac{d^2\psi(r)}{dr^2} + \frac{2\mu}{\hbar^2}\left[E - \left(\frac{A}{r} + \frac{B}{r^2}\right)e^{-\alpha r} - \frac{L\hbar^2\alpha^2}{2\mu(1-e^{-\alpha r})^2}\right]\psi(r) = 0, \qquad (3)$$

where, $L = l(l+1)$. To solve for the approximate solution, we use the approximation scheme of Greene and Aldrich [43] and the coordinate transformation of $z = e^{-\alpha r}$. We obtain the expression,

$$z(1-z)\frac{d^2\psi(z)}{dz^2} + \frac{1}{z(1-z)}\frac{d\psi(z)}{dz} + \frac{1}{z(1-z)}\left[-\beta(1-z)^2 + \frac{2\mu A}{\alpha\hbar^2}z(1-z) - \frac{8\mu B}{\hbar^2}z - L\right]\psi(z) = 0, \qquad (4)$$

where,

$$\beta = -\frac{2\mu E}{\alpha^2\hbar^2} \qquad (5)$$

Eqn (4) is further evaluated to obtain

$$z(1-z)\frac{d^2\psi(z)}{dz^2} + (1-z)\frac{d\psi(z)}{dz} + \frac{1}{z(1-z)}\left[-\left(\beta - \frac{4\alpha\mu A}{\hbar^2\alpha}\right)z^2 + \left(2\beta - \frac{4\alpha A}{\hbar^2\alpha} - \frac{8\mu B}{\hbar^2}\right)z - (\beta + L)\right]\psi(z) = 0 \qquad (6)$$

The hypergeometric equation obtained as,

$$\frac{d^2\psi(z)}{dz^2} + \frac{1}{z}\frac{d\psi(z)}{dz} + \frac{1}{z^2(1-z)^2}\left[-\chi z^2 + \Lambda z - \Sigma\right]\psi(z) = 0 \qquad (7)$$

where we have introduced the following notations for mathematical simplicity;

$$\beta + \frac{4\alpha\mu A}{\hbar^2\alpha} = \chi \qquad (8)$$

$$2\beta - \frac{4\alpha A}{\hbar^2\alpha} - \frac{8\mu B}{\hbar^2} = \Lambda \qquad (9)$$

$$\beta + L = \Sigma \qquad (10)$$

Eqn (7) is solved by factorization method [44] and the energy eigenvalue expression is obtained as,

$$E_{nl} = -\frac{\alpha^2 \hbar^2}{2\mu} \left\{ \left( \frac{\frac{2\mu A}{\alpha \hbar^2} + l(l+1) + (\omega+n)^2}{2(\omega+n)} \right)^2 + l(l+1) \right\} \qquad (11)$$

The corresponding expression for the wavefunction is obtained as,

$$\psi_{n\ell}(z) = N_n z^{\mu} (1-z)^{\omega} {}_2F_1\left(-n, n+2(\mu+\omega); 1+2\mu; z\right), \qquad (12)$$

where, $z = e^{-\alpha r}$, $\mu = -\frac{2\mu E}{\alpha^2 \hbar^2} + l(l+1)$, $\omega = \frac{1}{2} + \sqrt{\frac{2\mu B}{\alpha^2 \hbar^2} + \left(l+\frac{1}{2}\right)^2}$, and $N_n$ is the normalization constant and $\mu$ and $\hbar$ are in atomic units.

## 3. Thermodynamic properties of the modified Dirac delta distribution

In term of the modified Dirac delta statistical model, the Boltzmann factor is defined as [38],

$$B(E) = e^{-\beta E}\left(1 + \frac{q}{2}\beta^2 E^2\right) \qquad (13)$$

where $q$ is defined as the deformed parameter and lies between zero and unity. For $q \to 0$, (9) reduces to normal statistics. The parameter $\beta$ is the fluctuation parameter and it is an inverse relation of the Boltzmann constant, $K_b$ is taken as unity, and temperature T in Kelvin as,

$$\beta = \frac{1}{K_b T} \qquad (14)$$

To examine the thermodynamic function for the deformed statistics, we would obtain the partition function. The partition function is defined as,

$$Z(\beta) = \int B(E) = \int e^{-\beta E}\left(1 + \frac{q}{2}\beta^2 E^2\right) dn \qquad (15)$$

Substituting eqn (8) into (11), we obtain the expression,

$$Z(\beta) = \frac{q\sqrt{-\beta\lambda}\ e^{y_1^2\beta\lambda}}{8(-\beta\lambda)^{\frac{3}{2}}} \left( \frac{2y1^3\beta\lambda + 2y1^2\beta\lambda\sqrt{-P1+y1^2} + y1(-3+4P2\beta\lambda) +}{\sqrt{-P1+y1^2}\ (-3+2\beta\lambda P1 + 4\beta\lambda P2)} \right) -$$

$$\frac{q\sqrt{-\pi\beta\lambda}\ e^{y_1^2\beta\lambda}}{8(-\beta\lambda)^{\frac{3}{2}}} \left( 8 + q \left( \begin{array}{c} 3 - 4P2\beta\lambda + (2P2\beta\lambda)^2 + \\ e^{P1\beta\lambda}\left(8 + q\left(3 + 4\beta\lambda(P1+P2)(-1+(P1+p2)\beta\lambda)\right)\right) \end{array} \right) \right) -$$

$$\frac{q\beta\lambda\ e^{y_1^2\beta\lambda}}{4(-\beta\lambda)^{\frac{3}{2}}} \left( 8 + q\left(3 - 4P2\beta\lambda + (2P2\beta\lambda)^2\right) \right) \text{DawsonF}\left[y1\sqrt{\beta\lambda}\right] +$$

$$\left( 8 + q\left(3 + 4(\beta\lambda)^2(P1+P2)(-1+(P1+P2))\right) \right) \text{DawsonF}\left[\sqrt{(-P1+y1^2)(\beta\lambda)}\right]$$

(16)

where,

$$P1 = \frac{2\mu A}{\alpha^2\hbar^2} + l(l+1) ,\tag{17}$$

$$P2 = l(l+1),\tag{18}$$

$$\lambda = \frac{\hbar^2\alpha^2}{2\mu},\tag{19}$$

$$y1 = \frac{P1}{1+2\sqrt{\frac{2\mu B}{\hbar^2\alpha^2}+\left(l+\frac{1}{2}\right)^2}} + \left(\frac{1}{4} + \frac{1}{2}\sqrt{\frac{2\mu B}{\hbar^2\alpha^2}+\left(l+\frac{1}{2}\right)^2}\right)\tag{20}$$

The DawsonF is the Dawson integral given by

$$\text{DawsonF} = F(x) = e^{-x^2}\int_0^x e^{y^2} dy \tag{21}$$

4. **Thermodynamic properties for normal distribution**

The uniform distribution for a parameter $\beta > 0$ is defined as [38]

$$f(\beta) = \frac{1}{b}, \tag{22}$$

where $\beta$ is within the interval $[x, x+y]$ and zero elsewhere. In terms of Boltzmann factor is expressed as [58,45],

$$B_b = \int_0^\infty e^{-\beta E} f(\beta) d\beta \approx e^{-\beta E} \left[1 + \frac{1}{24} b^2 E^2 \right], \tag{23}$$

where, $b \to 0$ reduces to the normal statistics. The partition function for uniform distribution is,

$$Z_b(\beta) = \int B_b(E) dn \tag{24}$$

The thermodynamic properties in terms of partition function $Z(\beta)$ are,

Free Energy, F

$$F = \frac{\ln Z}{\beta} \tag{25}$$

Mean Energy, U

$$U = -\frac{\partial \ln Z}{\partial \beta} \tag{26}$$

Entropy, S

$$k \ln z - k\beta \frac{\partial \ln Z}{\partial \beta} \tag{27}$$

Specific Heat, C

$$C = k\beta^2 \frac{\partial^2 \ln Z}{\partial \beta^2} \tag{28}$$

The thermodynamic function of the deformed distribution and normal distribution will be evaluated from eqns (25) – (28).

## 5. Applications

We examine the thermodynamic properties for the diatomic molecules of $LiH$, $HCl$ and $H_2$. We consider the spectroscopic parameters for SKP in (1) [42, 46] for $\hbar = 0.0197329$ as shown in the table below.

Table 1: Spectroscopic parameter of some selected diatomic molecules

| Molecules | $D_e(eV)$ | $r_e(A°)$ | $\mu(a.m.u)$ |
|---|---|---|---|
| $HCl$ | 4.619030905 | 1.2746 | 0.09129614886 |
| $LiH$ | 2.5152672118 | 1.5956 | 0.08198284801 |
| $H_2$ | 4.7446 | 0.7416 | 0.04693891556 |

We evaluated for the partition function using the energy eigenvalue obtained SKP in eqn (11) for $l = 0$. The expression is presented in eqn (16). The partition function obtained was used for the analysis of thermodynamic properties as seen in eqns (25) through (28) for the diatomic molecules of $LiH$, $HCl$, and $H_2$ in terms of the q-deformed distribution and uniform distribution for various values of $q$ and $b$, where the values of $q$ and $b$ lies between zero and unity. We performed graphical analysis to observe their comparative behavior in the absence $(q = b = 0)$ and as well in the presence of $q$ and $b$ parameters. Fig 2 are various behaviors of the partition function of the of $LiH$, $HCl$ and $H_2$ molecules. The diatomic molecules were observed to possess similar characteristics in the presence and absence of the q deformed parameter. As value of q increased, the partition function showed and inverse relation to $\beta$, the inverse temperature. It was also observed that in the presence of $q$, as it increases it moves downward. In fig 3, we studied the behavior the diatomic molecules would exhibit for free energy and we observed a gradual exponential decay along $\beta$ for the various values of $q$. In the presence of $q$, as the values of $q$ is increased, it moves upward. Fig 4, is the entropy behavior for the various diatomic molecules. In the absence and presence of $q$ show a striking dissimilarities. For $q = 0$ what occurred was a monotonic decreases along $\beta$. In the presence of $q$, there was a steep decrease and a gradual rise occurred afterwards. This changes occurred between $\beta = 0.2$ and $\beta = 0.4$ for the different diatomic molecules. Fig 5 and 6 are the plots of the mean energy and specific capacity for the various diatomic molecules. Both thermodynamic function took to an exponential increasing manner and in the absence and presence of $q$. For the increasing values of $q$, it took to an steeper exponential rise. Figures 7 to 10 are the various plots of the uniform distribution for the various thermodynamic properties for the diatomic molecules of $LiH$, $HCl$, and $H_2$. We plotted for the partition function in fig. 7, unlike the deformed counterpart in fig 2, the decrease was not linear but the representation shows a monotonic decreasing partition function in the forward direction. The free energy plot in fig. 8, as it is it in free energy in modified Dirac delta distribution of fig. 3, so it is in uniform distribution as their behavior exhibited similar characteristics and the absence and presence of $b$. At $b=0$ in fig 9, for an increasing entropy function, it witnessed a gradual exponential decay, however, in the presence of $b$, as the entropy increased gradually and then a steep rise. In fig 10, and 11, the mean energy and specific capacity respectively, showed a similar increasing gradient as the values of $b$ increased.

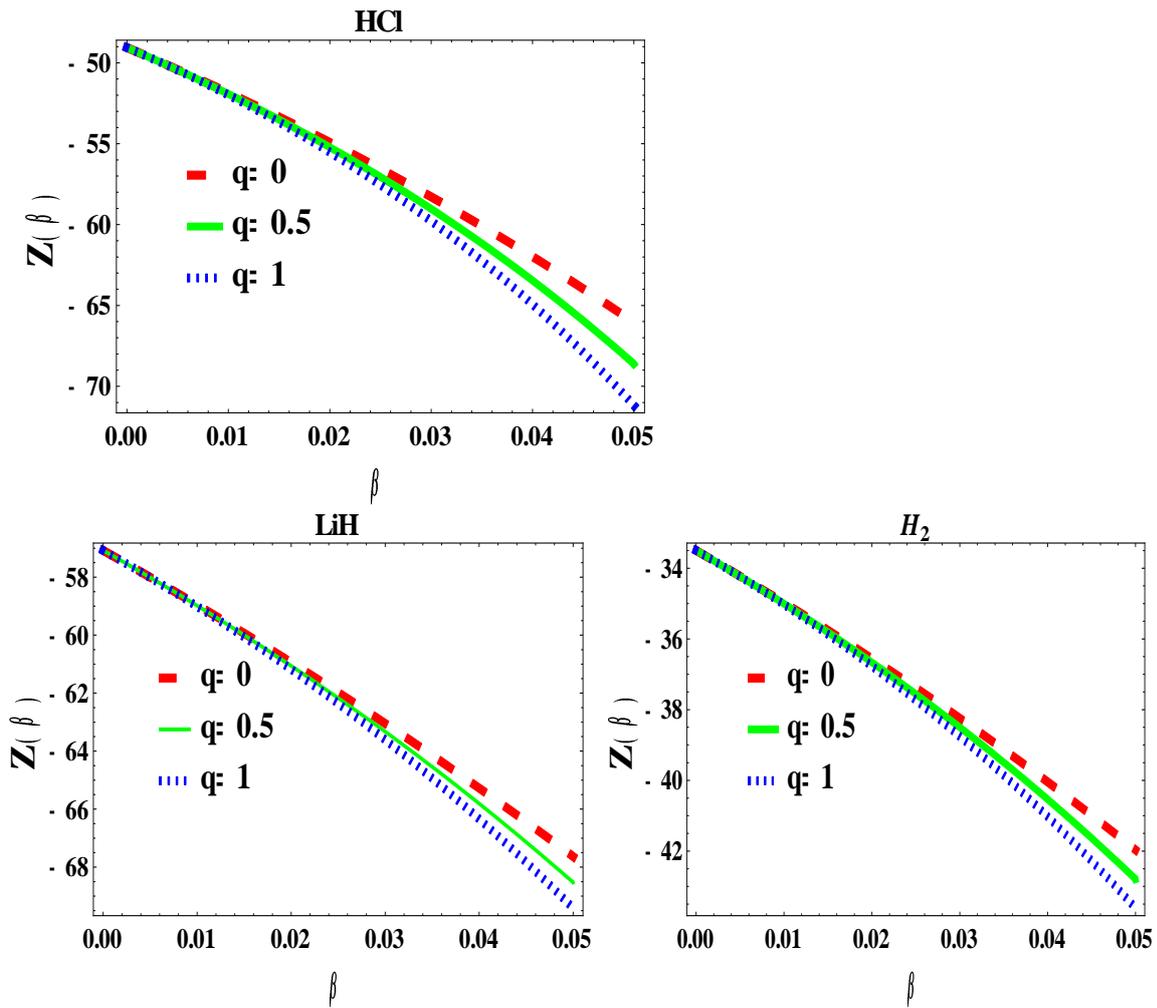

Fig 2. Partition function for the q deformed distribution for the diatomic molecules of *LiH*, *HCl* and $H_2$

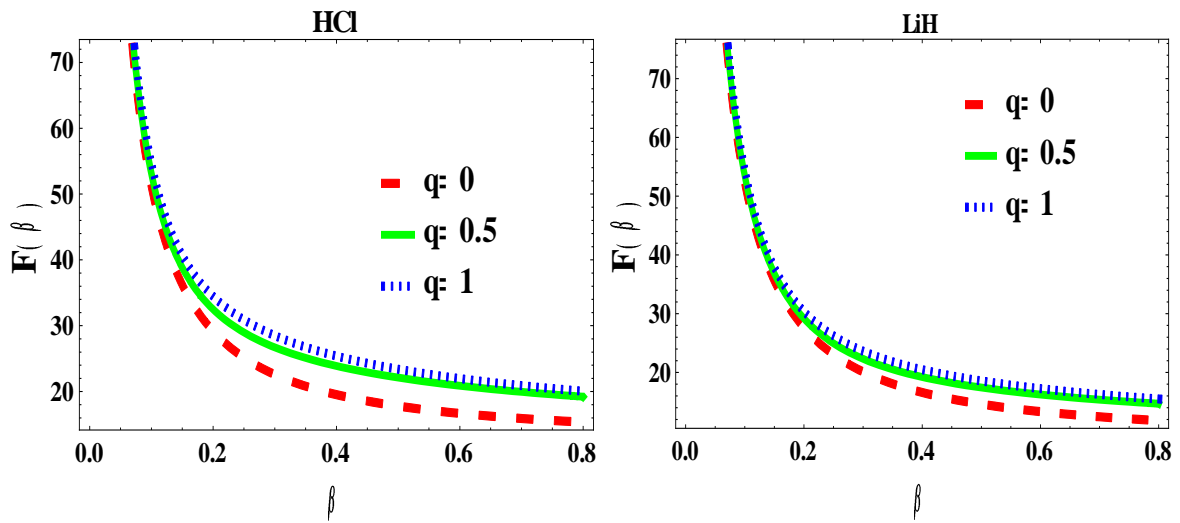

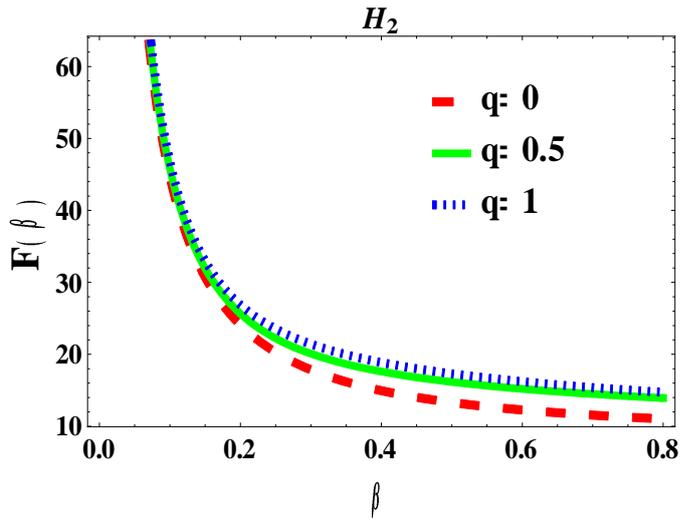

Fig 3. Free energy for the diatomic molecules for the q deformed distribution for the various diatomic molecules of *LiH* , *HCl* and $H_2$

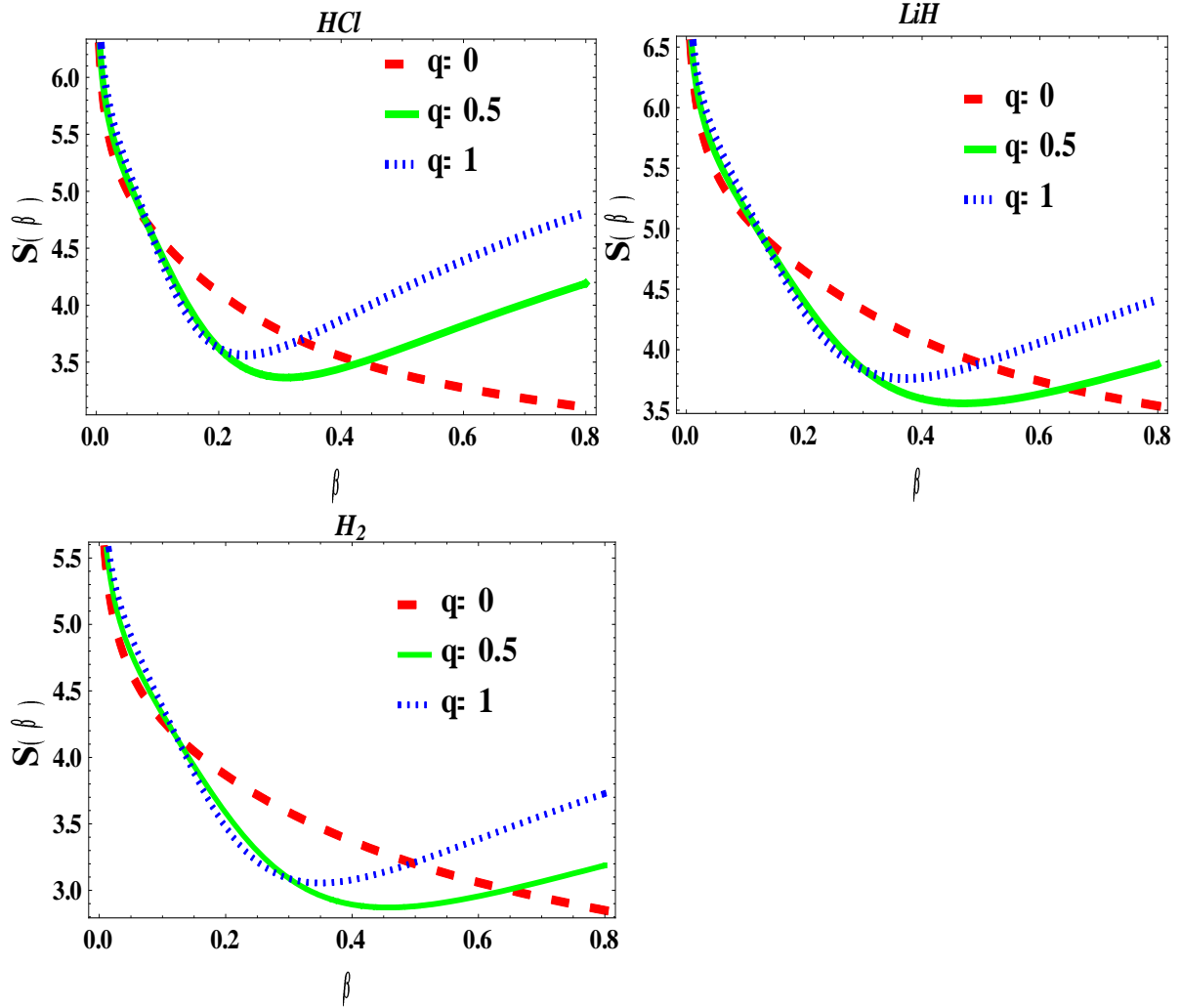

Fig.4. Entropy of the q deformed distribution for the various diatomic molecule of *LiH* , *HCl* , and $H_2$

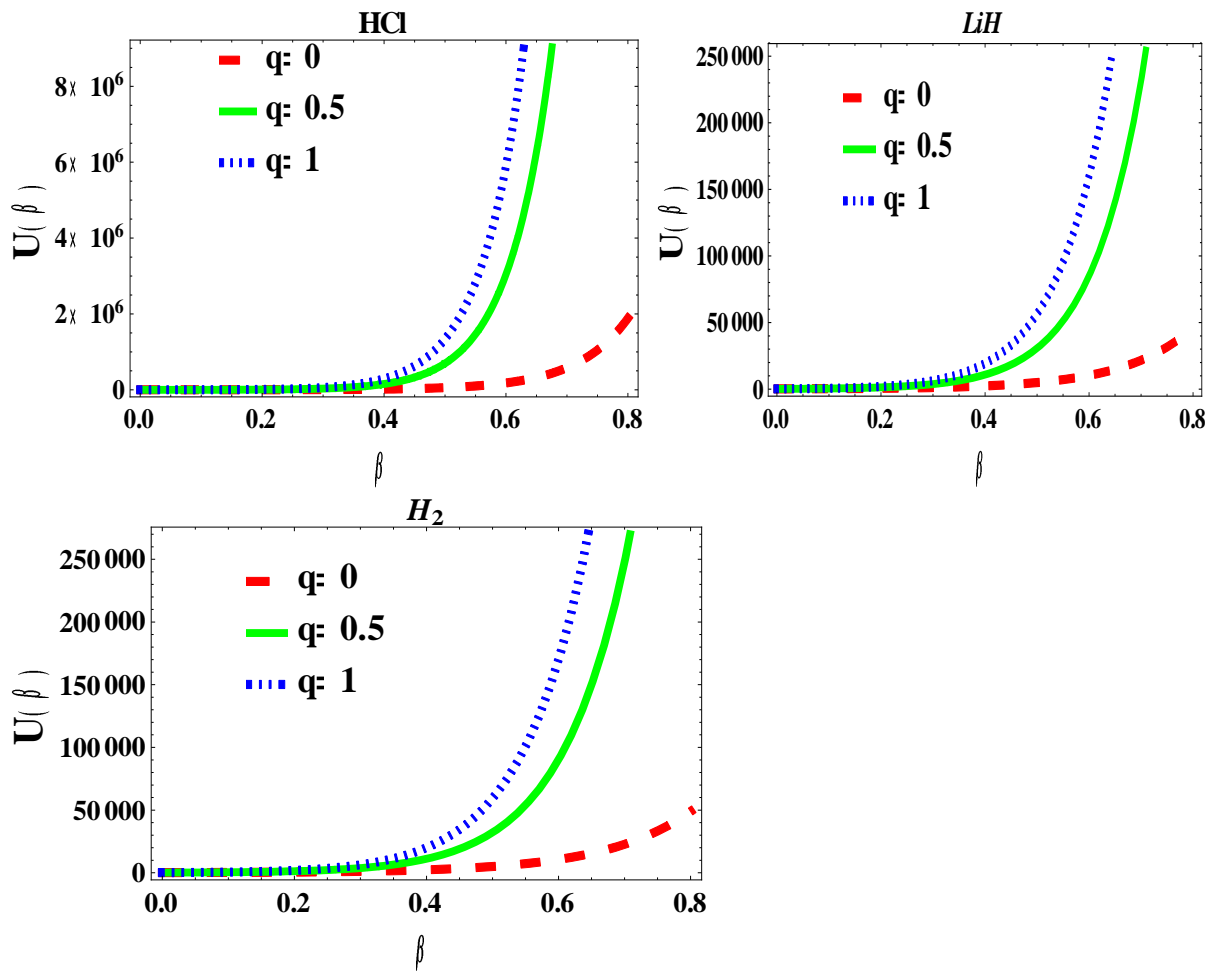

Fig. 5. Mean energy of the q deformed distribution for the diatomic molecules of $LiH$, $HCl$, and $H_2$

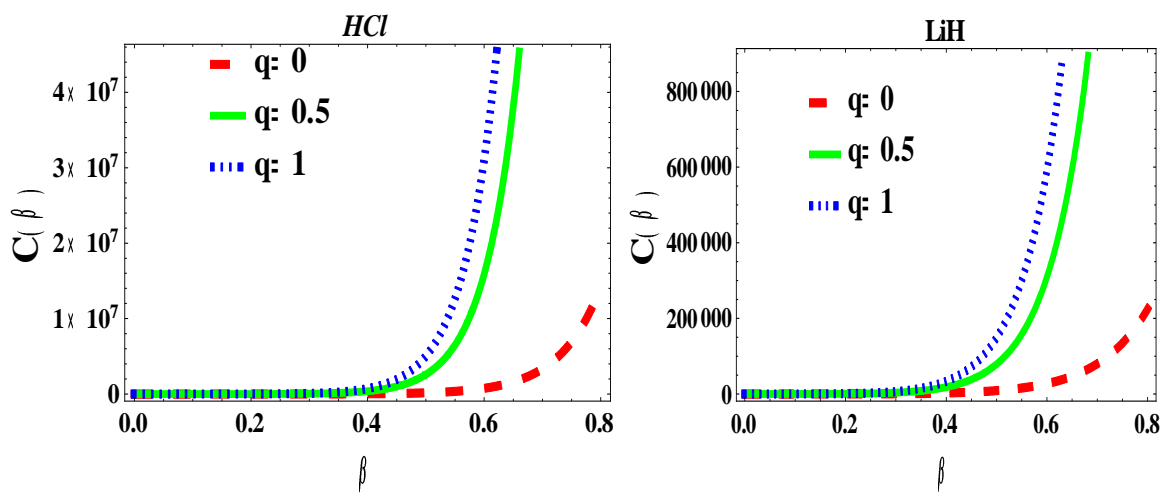

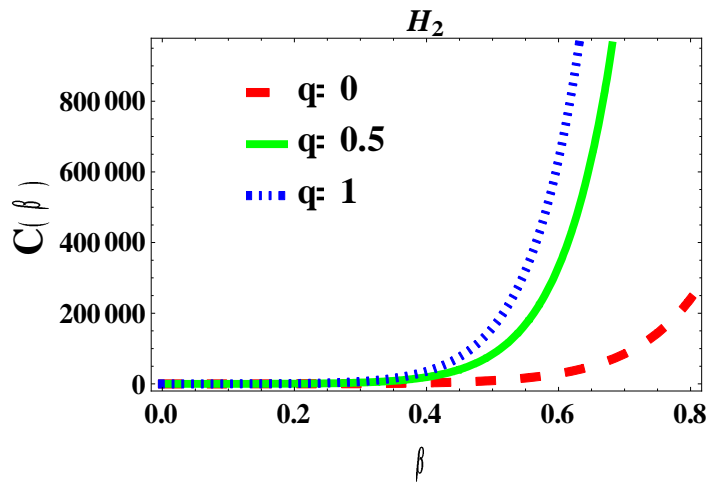

Fig. 6. Specific capacity of the q deformed distribution for the diatomic molecules of $LiH$, $HCl$ and $H_2$

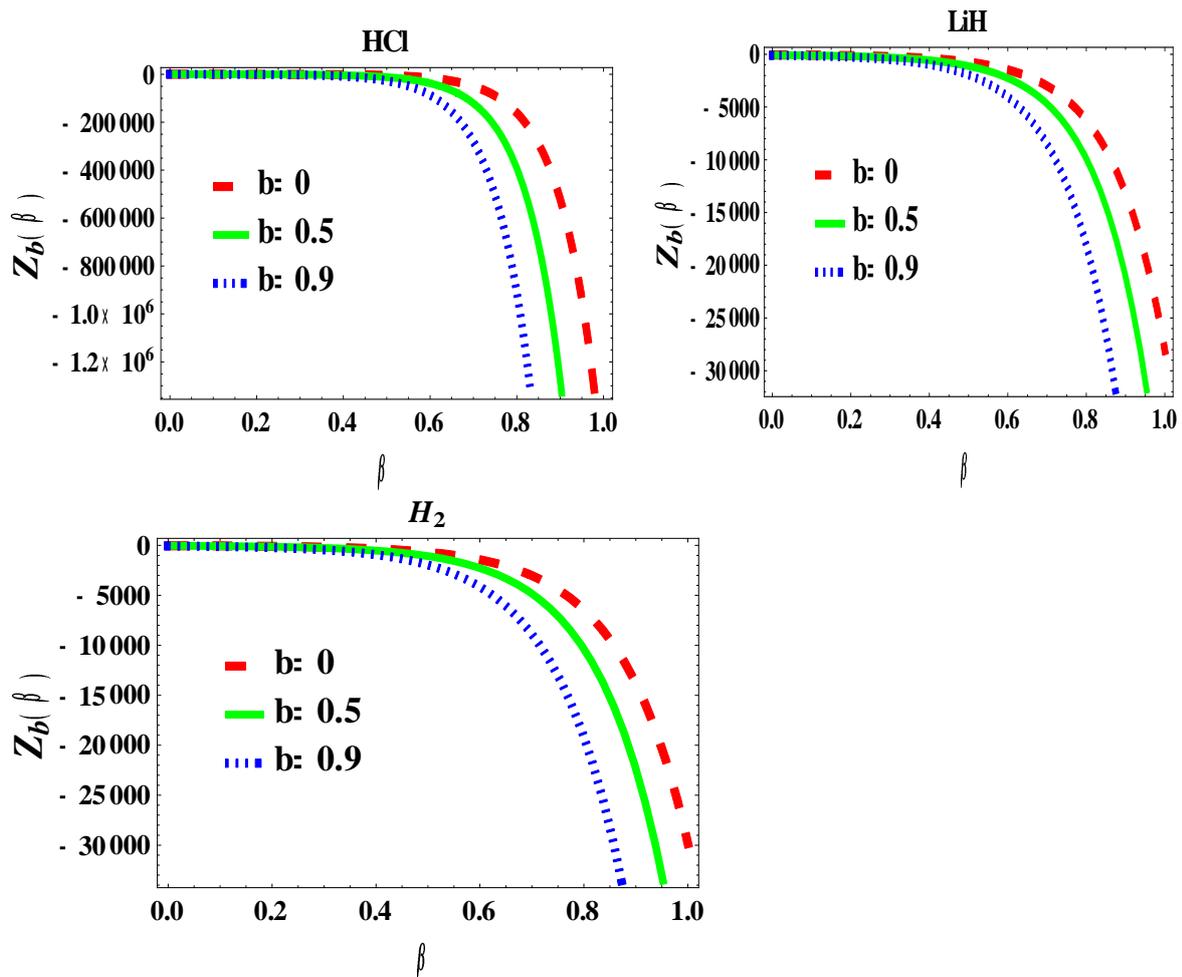

Fig. 7. Partition function of the uniform distribution for the diatomic molecules of of $LiH$, $HCl$ and $H_2$

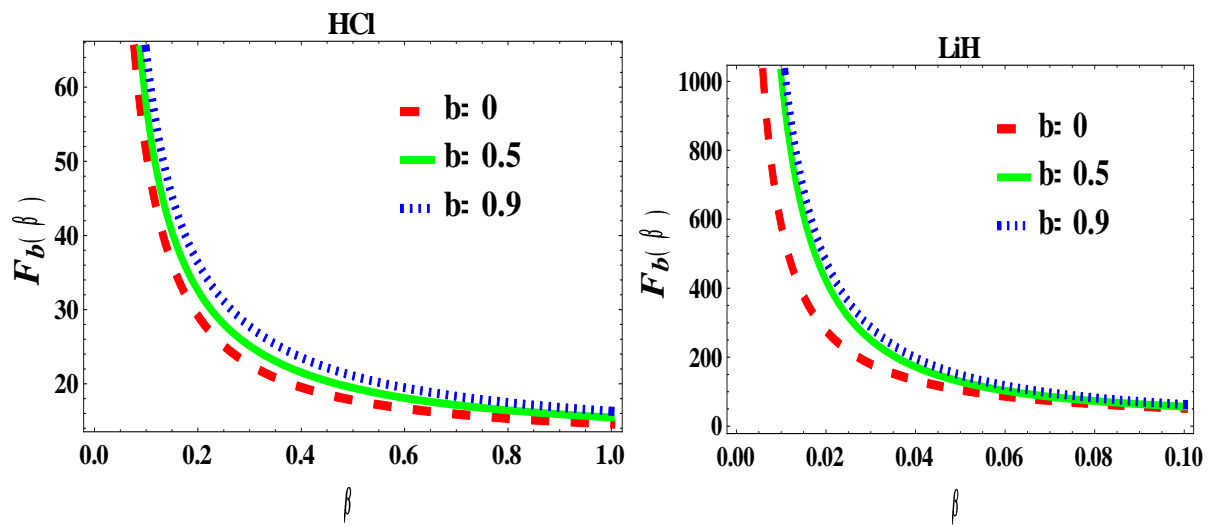

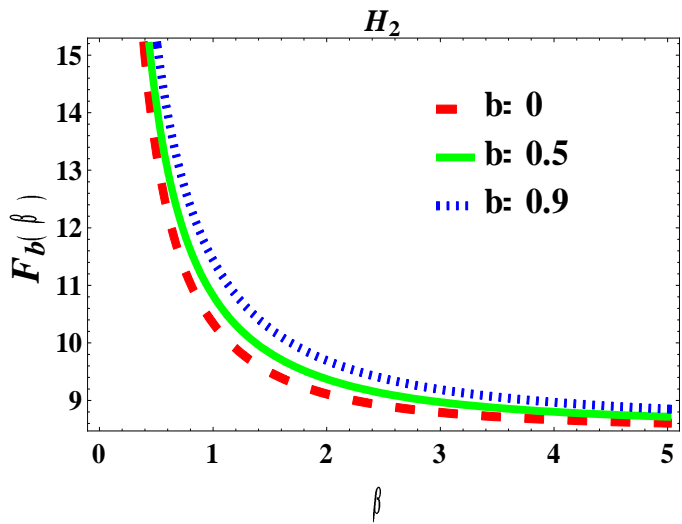

Fig. 8. Free energy of uniform distribution for the diatomic molecules of *LiH*, *HCl* and $H_2$

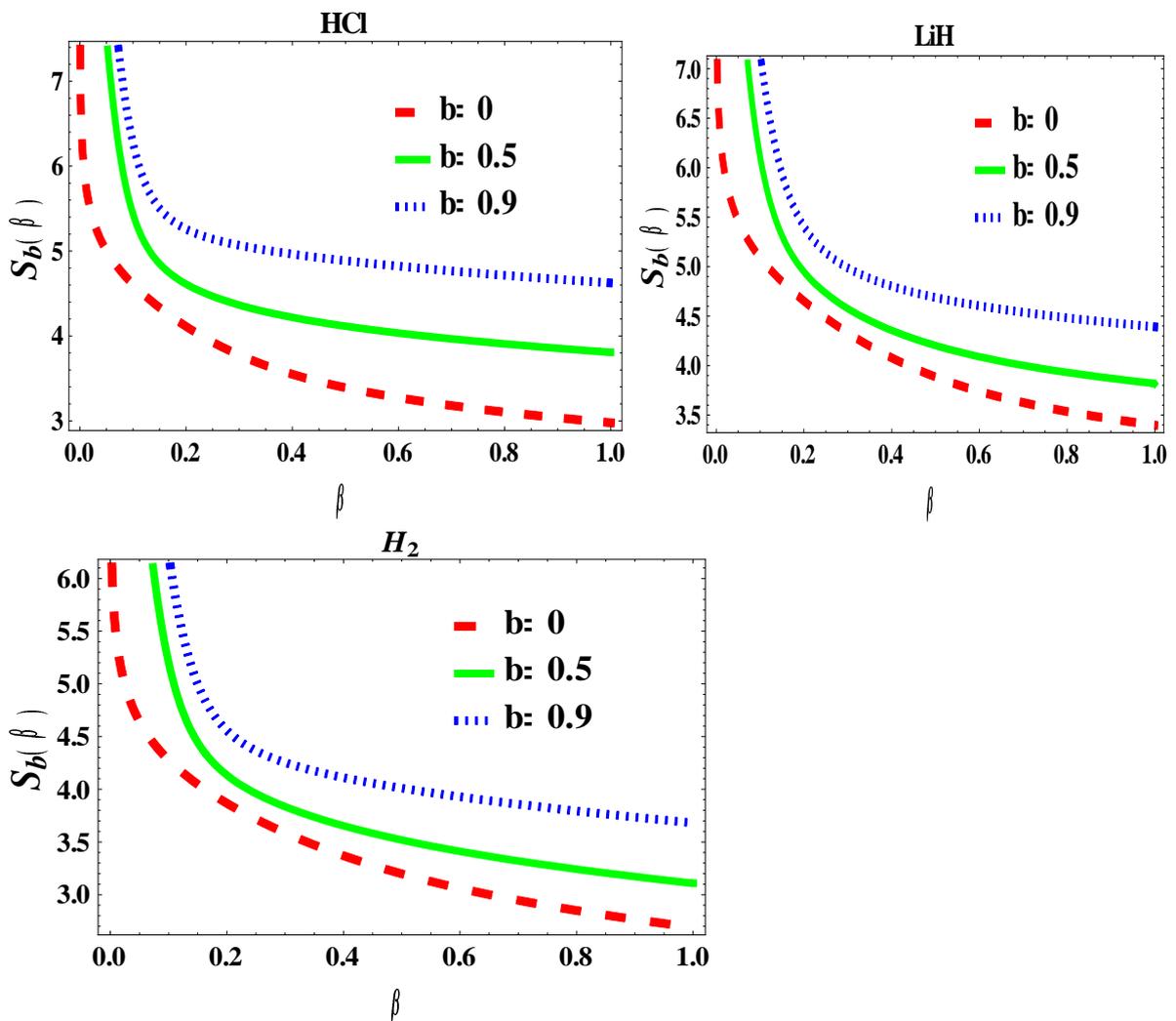

Fig 9. Entropy of the uniform distribution for the various diatomic molecules of *LiH*, *HCl* and $H_2$

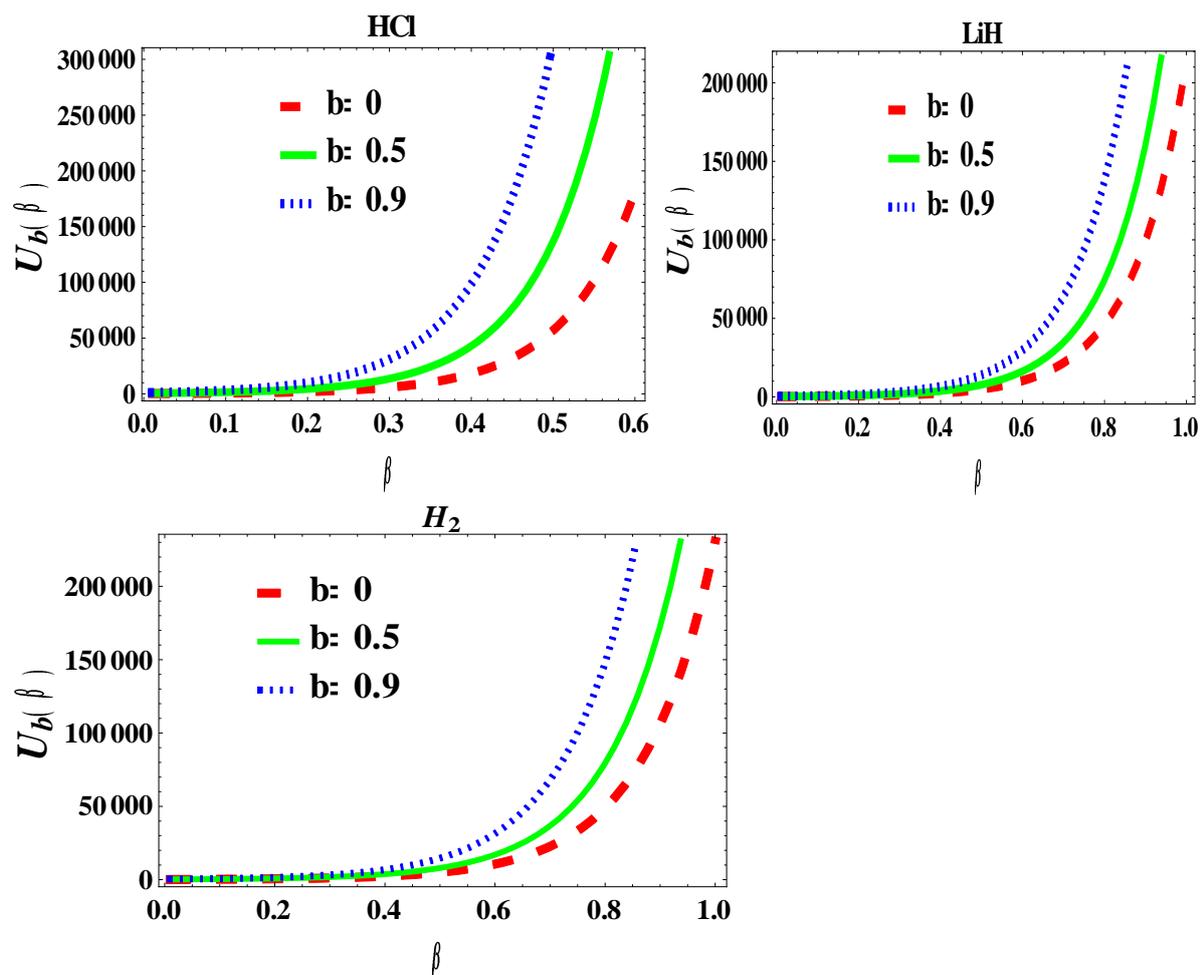

Fig 10. Mean energy of uniform distribution for the diatomic molecules of *LiH* , *HCl* and $H_2$

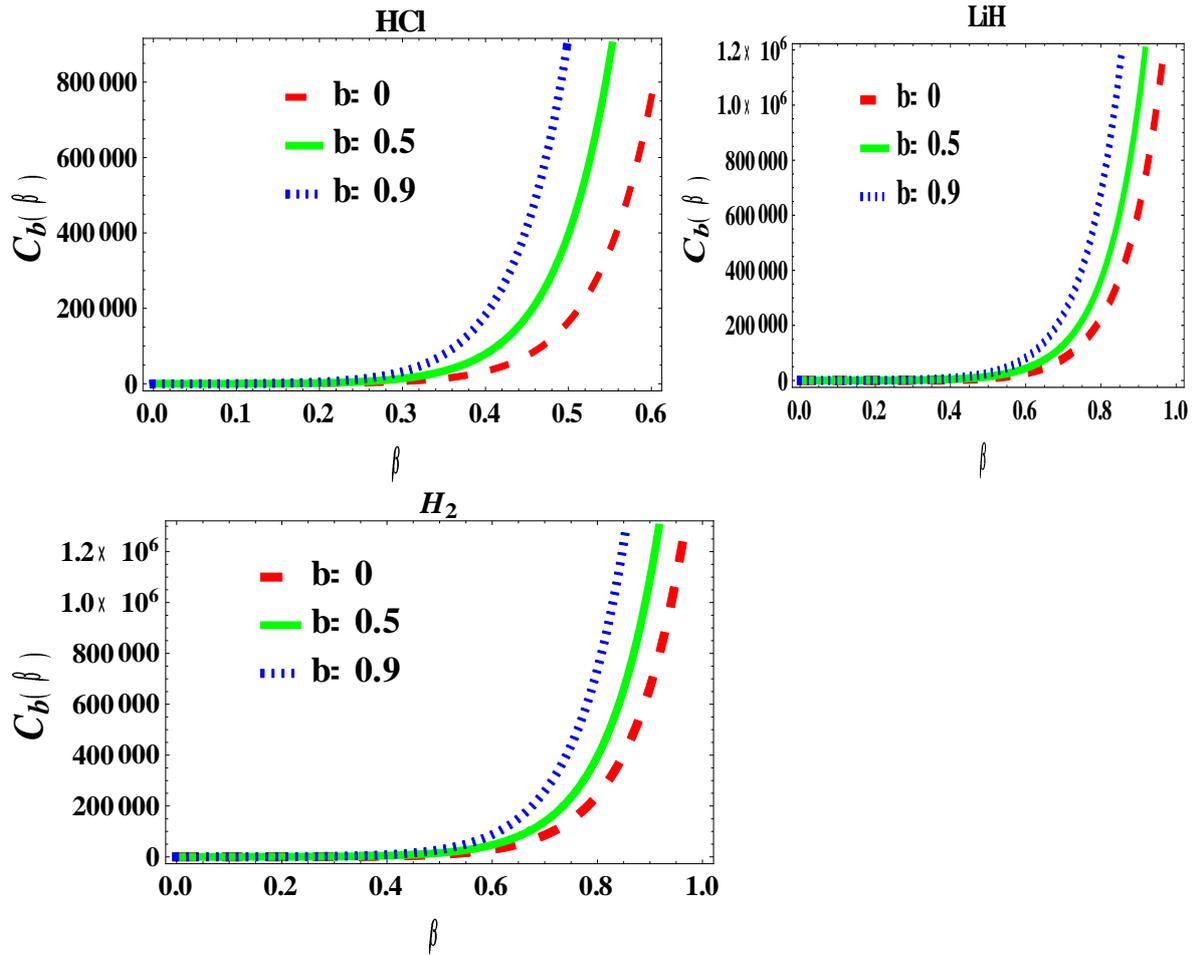

Fig 11. Specific capacity of uniform distribution for the diatomic molecules of $LiH$, $HCl$ and $H_2$

## 6. Conclusion

In this work, the eigenvalues obtained for the for SKP to evaluate for the partition function within the framework of the deformed formalism. The result obtained was used for examine thermodynamic properties for the modified Dirac delta and the uniform distribution for the diatomic molecules of $LiH$, $HCl$ and $H_2$. We have studied their behaviors in the absence (normal distribution) and also in the presence of the $q \text{ and } b$ parameters and observed them comparatively for the different thermodynamic function.